# Structure–Function Coherent Coarsening for Cross-Resolution Ecohydrological Modeling

Long Jiang[1,2,*], Yang Yang[3], Morgan Thornwell[4], Robert McKane[5], Sonali Chokshi[6], Huanfeng Duan[2], Tiantian Yang[7], Hoshin Vijai Gupta[8]

[1]*Puget Sound Institute, University of Washington, Tacoma, Washington, USA*

[2]*State Key Laboratory of Climate Resilience for Coastal Cities, Department of Civil and Environmental Engineering, The Hong Kong Polytechnic University, Hong Kong SAR, China*

[3]*School for the Environment, University of Massachusetts Boston, Boston, Massachusetts, USA*

[4]*Thornwell Labs, Portland, Oregon, USA*

[5]*Healthy Watersheds LLC, Philomath, OR, USA*

[6]*Oak Ridge Associated Universities, Corvallis, OR, USA*

[7]*School for Environment and Sustainability, University of Michigan, Ann Arbor, Michigan, USA*

[8]*Department of Hydrology and Atmospheric Sciences, The University of Arizona, Tucson, Arizona, USA*

**Email: jlon@uw.edu*

## Key Points

- A structure–function coherent coarsening framework preserves drainage topology and land–soil functional heterogeneity across spatial scales.
- Hydro-Aware DEM coarsening and class-adaptive aggregation support physically consistent and stable runoff and nitrogen-loss simulations.
- The framework can reduce computational costs for calibration and scenario analysis, support multi-source integration, and improve transferability.

## Abstract (249 words)

Ecohydrological models are increasingly applied in regional scenario-based studies, yet their cross-resolution application remains constrained by high computational costs of fine-resolution simulations and by structural inconsistencies introduced during scale transfer. We propose a *Structure–Function Coherent Coarsening* (SFCC) framework that improves the preservation of hydrological structure and functional heterogeneity during model input coarsening.

Using the VELMA model in 24 subbasins of the Salish Sea Basin, USA, we evaluate coarsening strategies for three input types: (i) DEMs coarsened with a *Hydro-Aware* approach that better preserves drainage topology; (ii) land cover and soil type datasets coarsened with function-preserving methods (*Auto-Weight* and *Auto-Reassign*) that retain small but process-dominant classes; and (iii) initial conditions coarsened with *Hydro-Aware*, *Landcover-Aware*, and *Soil-Aware* strategies to improve temporal stability.

Results show that the *Hydro-Aware* method better preserves watershed morphology and yields more consistent runoff and nitrate loss predictions across resolutions than mean-based coarsening. For categorical inputs, the function-preserving methods can reduce the dominant-class bias of majority aggregation, particularly in basins where small, high-impact patches drive nitrogen export. Long-term simulations show that hydrological variables stabilize quickly, whereas biogeochemical variables stabilize more slowly. In both cases, differences caused by coarsening decrease over time and eventually level off. This suggests that preserving both structural consistency and functional heterogeneity can help maintain stable model behavior over time.

By operating directly at the data-input level, SFCC improves integration of multi-source datasets while retaining high-resolution information. It supports cross-scale simulation, calibration, and model integration, and provides a scalable foundation for regional scenario analysis.

## 1 Introduction (1321 words)

[1] In recent years, ecohydrological models have become important tools for assessing the potential impacts of land-use change, extreme events, and management interventions under intensifying climate pressures and human activity *(John et al., 2021; Keller et al., 2023; Silva et al., 2024)*. As applications expand, models are increasingly required to operate across multiple spatial resolutions and scenarios to support large-scale ecohydrological assessments. Higher spatial resolution better captures system dynamics, including slope runoff, soil heterogeneity, and nutrient cycling *(Jin et al., 2019; van Jaarsveld et al., 2025)*. However, fine-scale modeling also leads to a dramatic increase in computational cost, particularly during parameter calibration, where numerous iterative simulations are needed to identify acceptable parameter sets, posing a major bottleneck for large-scale and scenario-based modeling *(Herrera et al., 2022; Sun et al., 2020; Wu et al., 2025)*.

[2] To reduce computational cost while maintaining fidelity to fine-resolution processes, spatial coarsening of the input data (including digital elevation model (DEM), land cover, and soil type) has emerged as an effective strategy *(Armstrong and Martz, 2008; Von Gunten et al., 2014)*. At the same time, coarsening is also often used to reconcile multi-source datasets with inconsistent spatial resolutions, allowing high-resolution inputs to be consistently integrated into large-scale ecohydrological models. These low-resolution models can accelerate parameter calibration and simulations with limited loss of accuracy. Yet, the coarsening process can distort the physical and functional integrity of the data, altering hydrological and biogeochemical responses *(Son et al., 2016)*. These distortions not only undermine model stability and transferability, but also introduce structural biases across scales, which further lead to inconsistencies in the physical interpretation of parameters and process representation *(Ichiba et al., 2018; Singh and Kumar, 2017; Tudaji et al., 2025; Yang et al., 2019)*. Such inconsistencies challenge the ability of low-resolution models to reliably reproduce high-resolution process dynamics, thereby limiting their usefulness for effective parameter calibration and multi-scale simulation *(Merz et al., 2009; Moges et al., 2023)*.

[3] Early studies noted that parameters often exhibit systematic biases during aggregation or scale transitions, causing their "*effective meaning*" to drift with scale *(Blöschl and Sivapalan, 1995)*. In soil hydrology, point-scale parameters cannot be directly transferred to grid or watershed scales without constructing equivalent parameters *(Vereecken et al., 2007)*. The effective transfer of parameters from the point scale to the grid or watershed scale relies heavily on the assumptions of homogeneity and representative elementary volumes, which rarely hold in heterogeneous landscapes. Furthermore, parameter non-uniqueness and compensatory effects (equifinality) limit the robustness of cross-scale transfer *(Beven, 2006)*. To address these issues, the multiscale parameter regionalization (MPR) framework *(Kumar et al., 2013; Samaniego et al., 2010)* introduced transfer functions and aggregation operators to upscale high-resolution environmental variables into effective coarse-scale parameter values, enhancing regional consistency. However, its performance depends heavily on high-quality priors and deteriorates under non-stationary or extrapolative conditions *(Rakovec et al., 2016)*. Hierarchical Bayesian and surrogate modeling approaches provide alternative approaches for multi-scale calibration *(Asher et al., 2015)*, but remain computationally intensive and may lack physical interpretability.

[4] The SUMMA framework *(Clark et al., 2015)* explicitly decouples structural and parametric uncertainties, revealing how the physical interpretation of parameters can systematically shift with model structure and spatial resolution. This finding demonstrates that recalibration alone cannot eliminate scale-induced structure–parameter coupling. Similarly, the "hyper-resolution modeling" paradigm does not necessarily improve predictive reliability *(Beven and Cloke, 2012; Wood et al., 2011)* and that mismatched data accuracy and process representation can instead amplify model uncertainty. Collectively, these insights suggest that coarsening should not be treated as a mere reduction in resolution, but as a physically constrained process that must preserve both hydrological structure and functional representativeness. Without such constraints,

computational costs and uncertainties rise in tandem *(Wu et al., 2025)*, undermining cross-scale consistency and model transferability across applications.

[5] In general, existing coarsening methods for ecohydrological models fall into two categories: (i) statistical coarsening and (ii) structure-preserving coarsening. For continuous variables such as DEMs, statistical coarsening aggregates grid cells using the mean, median, maximum, or minimum operators. Although computationally efficient, this approach smooths terrain, alters slopes and drainage patterns, and introduces significant deviations once the resolution falls below a critical threshold *(Aristizabal et al., 2024; Hoang et al., 2018; Ichiba et al., 2018; Mahmood and Vivoni, 2011)*. By contrast, structure-preserving coarsening explicitly incorporates flow direction, accumulation, stream masks, and hydro-conditioning techniques such as stream burning and pit filling to maintain topological integrity *(Chen et al., 2024; Jiang et al., 2023; Moges et al., 2023)*. These methods improve drainage delineation and hydrological connectivity *(Moretti and Orlandini, 2018)* but remain computationally demanding and lack standardized optimization criteria and validation benchmarks. For discrete categorical inputs such as land cover and soil type, coarsening presents distinct challenges because class-based data cannot be meaningfully averaged or interpolated. In practice, the majority aggregation method remains the most common approach. While it preserves the dominant class, it often distorts class proportions, obscures small but process-dominant units, and reduces predictive performance *(Armstrong and Martz, 2008; Fan et al., 2020; Shadmehri Toosi et al., 2025)*. Recent studies have further shown that even fine-scale soil structural heterogeneity can strongly influence infiltration–runoff partitioning *(Bonetti et al., 2021)*, highlighting the need to preserve both structural integrity and functional heterogeneity during spatial coarsening.

[6] Overall, significant challenges remain despite advances in DEM processing, categorical aggregation, and multi-scale modeling. Most existing approaches emphasize terrain structure and runoff, while overlooking the functional properties of land and soil inputs. How input-resolution degradation translates into output-performance loss is still unclear, and a unified coarsening framework that jointly preserves structural connectivity and functional representativeness is lacking. These gaps limit the stability and scalability of multi-scale ecohydrological modeling.

[7] Building on these insights, this study integrates structure-consistent and function-preserving coarsening approaches across different input types. We propose and systematically evaluate a *Structure–Function Coherent Coarsening* (SFCC) framework for multi-resolution ecohydrological modeling. Using 24 representative subbasins across the Salish Sea Basin, we apply the process-based VELMA (Visualizing Ecosystem Land Management Assessments) model to assess and compare the performance of various coarsening strategies:

- **DEM coarsening:** A *Hydro-aware* method weighted by flow accumulation to preserve terrain and hydrological connectivity.
- **Land cover and soil type coarsening:** The algorithms: *Auto-weight* (adaptive weighting by class importance) and *Auto-reassign* (dynamic redistribution based on distributional bias) methods to retain small but high-impact classes.
- **Continuous initial conditions:** The *Hydro-aware*, *Landcover-aware*, and *Soil-aware* methods to maintain consistency between initial states, flow networks, and ecological attributes.

[8] Importantly, this study is not framed as the development of a conventional resampling technique for spatial aggregation. Instead, the SFCC framework conceptualizes coarsening as a process-aware transformation, in which different input variables are aggregated according to their roles in governing hydrological and biogeochemical processes. By maintaining both the physical consistency of terrain–drainage structures and the functional representativeness of land–soil properties, the SFCC framework aims to enhance cross-scale model stability and realism while reducing computational cost. Acting directly at the data-input level, it alleviates structural inconsistencies that often arise during scale transitions. It also mitigates integration biases introduced by multi-source datasets with inconsistent spatial resolutions, thereby preserving key hydrological and ecological information. Moreover, SFCC is complementary to parameter-mapping approaches and multiscale

regionalization frameworks, providing a unified foundation for physically consistent and functionally coherent cross-scale ecohydrological modeling.

[9] This paper is organised as follows. Section 2 (Methods) introduces the study area and data sources, and describes the coarsening approaches and similarity metrics developed for three categories of inputs: the DEM, land cover and soil type, and continuous initial conditions. Section 3 (Results) systematically compares the effects of different coarsening strategies on flow structure, runoff, and nitrate loss simulations, revealing the scaling relationships between spatial resolution and model performance. Section 4 (Discussion) examines how the proposed structure–function coherence mechanism enhances cross-scale consistency and transferability, evaluates its potential for regional applications and model extension, and outlines possible directions for future research. Finally, Section 5 (Conclusion) summarizes the main findings and proposes a practical framework and outlook for large-scale, multi-scenario ecohydrological modeling.

## 2 Methods (1555 words)

### 2.1 Study Area and Data Sources

[10] This study focuses on the Salish Sea Basin in Washington State, USA. The basin encompasses the Puget Sound, the Strait of Juan de Fuca, and their adjacent coastal watersheds, covering a vast and topographically diverse region from coastal lowlands to mountainous uplands. This complex geomorphic setting represents the diversity and complexity typical of large-scale watershed systems. Within this domain, 24 major subbasins were selected for detailed analysis (**Fig. 1**). These subbasins exhibit a wide range of land cover types, including urban, agricultural, forested, and wetland landscapes. The diversity of the selected subbasins allows a systematic assessment of model responses under distinct ecohydrological process regimes. In addition, while the selected study region is representative of large-scale watershed systems across spatial scales, the region is of great importance for practical watershed planning and management. In summary, the selected study region provides an ideal testbed for cross-resolution model analysis and validation of our proposed SFCC framework within a grid-based ecohydrological modeling context.

[11] The VELMA ecohydrological model, developed by the U.S. Environmental Protection Agency (EPA), is a distributed, process-based ecohydrological model capable of simulating water balance, vegetation growth, and coupled carbon–nitrogen dynamics within spatially explicit grid cells. It has been widely applied to studies of watershed-scale water quality, hydrological fluxes, and ecosystem responses *(Abdelnour et al., 2013, 2011; Halama et al., 2023; McKane et al., 2025; Venable et al., 2025)*, making it particularly suitable for cross-scale scenario analyses and process-based investigations.

[12] The datasets used in this study encompass multiple environmental and hydrological components essential for the development and evaluation of VELMA ecohydrological model. Meteorological data, including precipitation and temperature time series, were obtained from the high-resolution PRISM dataset. Topographic information was derived from a 30 m DEM to support terrain and hydrological analyses. Land cover data were sourced from the National Land Cover Database (NLCD) and regionally adjusted to more accurately represent the land cover patterns across the Salish Sea Basin. Soil characteristics were extracted from the SSURGO/SOLUS databases, which provide detailed soil classifications and physicochemical properties. Streamflow calibration data were obtained from US Geological Survey (USGS) gauging stations, while nitrogen-related monitoring data were provided by the Washington State Department of Ecology. The specific sources of all datasets and software are provided in the Open Research section. Together, these diverse datasets form a comprehensive foundation for model parameterization, calibration, and cross-resolution evaluation.

[13] The models were calibrated across 24 major subbasins of the Salish Sea Basin for the period 2009–2019. The hydrological component was calibrated using daily streamflow observations from USGS gauging stations, focusing on key parameters controlling runoff generation, infiltration, and evapotranspiration. Parameter optimization was performed using a custom-developed reinforcement learning algorithm *(Morgan, 2025)* to reproduce both seasonal flow dynamics and long-term water balance. For the nitrogen cycle, monthly monitoring data from the Washington State Department of Ecology were used to calibrate key indicators,

including ammonium, nitrate, organic nitrogen, and organic carbon losses, with additional constraints from soil humus and detritus concentrations to ensure physical plausibility. The calibration aimed to reproduce observed seasonal variability and interannual fluctuations, accurately capturing nutrient cycling under different land cover and soil conditions. Parameter optimization employed an automated global Latin Hypercube Sampling (LHS) approach *(Jiang, 2025)*. Model performance was evaluated using the Nash–Sutcliffe Efficiency (NSE) metric, with mean NSE values exceeding 0.6 for streamflow and 0.3 for nitrogen-related variables, indicating good agreement with observations. Overall, these results demonstrate that the configured VELMA models exhibit stable structural behavior and process coherence across the Salish Sea Basin.

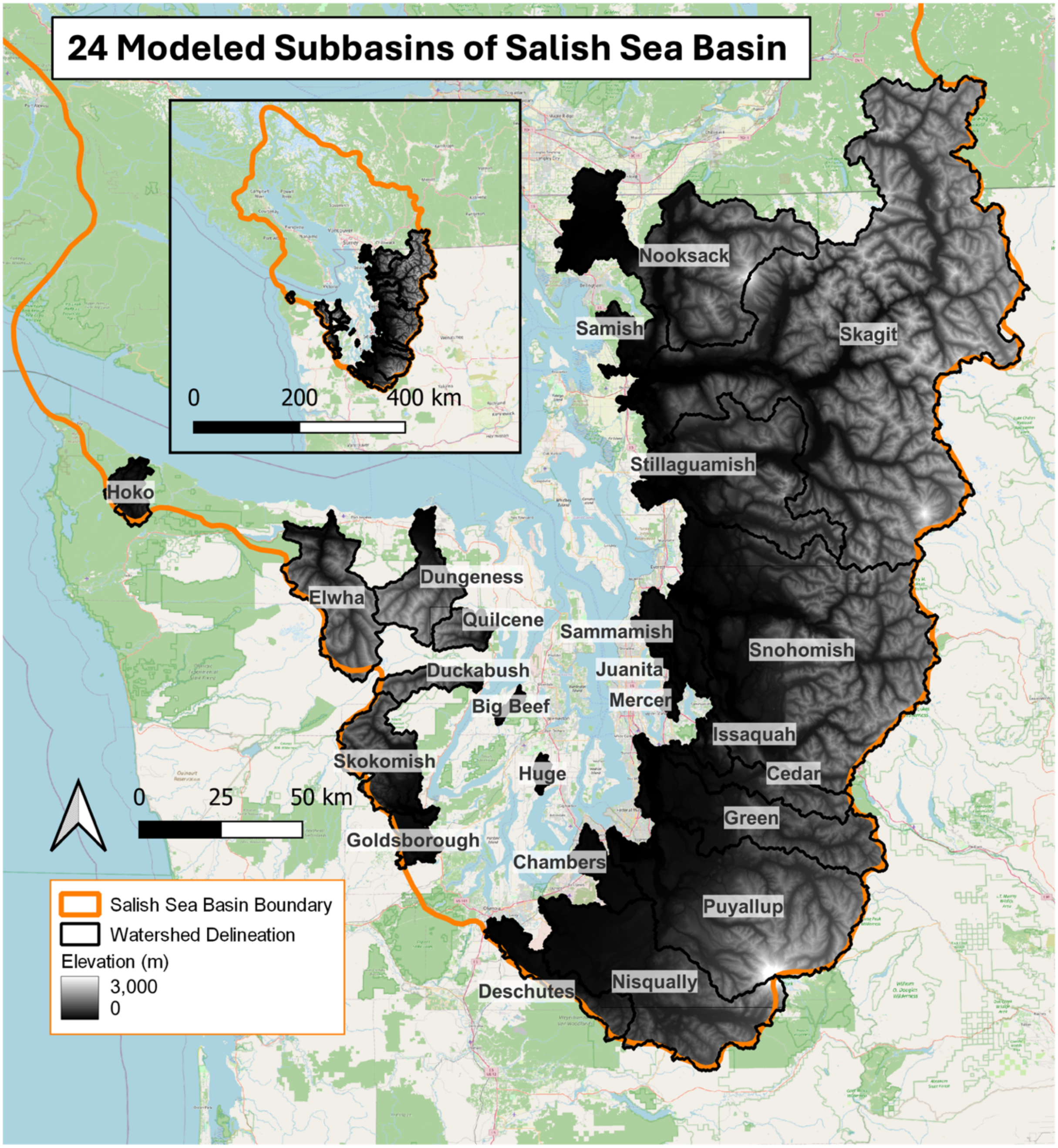

*Fig. 1 Spatial distribution of the 24 modeled subbasins within the Salish Sea Basin*

## 2.2 Coarsening Methods for Different Types of Input Data

[14] In VELMA, different ecohydrological processes exhibit distinct dependencies on spatial information. Runoff generation is simulated through a process-based water balance framework governed by precipitation, evapotranspiration, soil moisture storage, and lateral flow. Flow routing and accumulation are primarily controlled by topographic gradients and watershed topology, as represented by DEM-derived flow direction and connectivity. As a result, runoff behaves as a spatially continuous process that is highly sensitive to terrain information, with its accuracy depending on the preservation of flow paths and hydrological connectivity.

[15] In contrast, $NO_3$ loss is regulated by coupled biogeochemical processes, including mineralization, nitrification, plant uptake, and denitrification, with distinct parameterizations associated with different land cover and soil types. These processes are further coupled with hydrological transport via surface and subsurface pathways. Consequently, $NO_3$ dynamics are influenced not only by watershed topology but also strongly depend on land cover and soil type controls on nutrient transformation rates and retention capacity, with different categories contributing unequally to nitrogen cycling.

[16] Based on these differences, spatial coarsening should adopt differentiated, process-aware strategies for different types of input data. For DEM, preserving the underlying watershed topology is essential to maintain hydrological connectivity and flow routing. In contrast, for land cover and soil type data, it is critical to retain the functional contributions of different categories, particularly those with small spatial extent but disproportionate influence, to minimize process-level biases introduced by spatial aggregation.

[17] Guided by these principles, a suite of coarsening methods was proposed for different data types to reduce computational costs while preserving the spatial characteristics and physical consistency of model inputs (**Table 1** and **Fig. 2**). These approaches were specifically tailored to maintain spatial continuity, hydrological connectivity, and ecological representativeness across scales. **Table 1** summarizes the strategies applied to three major categories of input data: continuous variables such as DEM, categorical variables including land cover and soil type data, and continuous initial condition datasets such as nitrogen storage and soil moisture. Each method was designed to preserve the key structural or functional attributes most relevant to hydrological and biogeochemical processes. **Fig. 2** provides a brief illustration of the coarsening methods; the algorithmic implementation of all methods is available in the open-source toolkit accompanying this study *(Jiang, 2025)*. For clarity, Fig. 2 and Table 1 present simplified conceptual representations under a uniform grid assumption. More general formulations, including more detailed mathematical definitions and extensions to heterogeneous grids, are provided in the Supporting Information. Overall, these coarsening strategies are designed to align with the intrinsic characteristics of each type of data, providing a physically consistent and structurally robust foundation for cross-resolution ecohydrological modeling.

[18] It is worth noting that the robustness of flow direction and drainage topology across resolutions has been widely studied in terrain analysis, particularly through topology-preserving methods *(Liao et al., 2023; Tarboton, 1997; Yamazaki et al., 2017)*. In this context, rather than developing a new theoretical framework, the *Hydro-Aware* method serves as a representative and computationally efficient topology-preserving coarsening approach. In addition, the *Auto-Weight* and *Auto-Reassign* methods are not designed to reproduce the spatial distribution of high-resolution inputs alone, but to preserve spatial features that exert key controls on ecohydrological processes. Specifically, *Auto-Weight* enhances the influence of such critical units through aggregation weighting, whereas *Auto-Reassign* maintains their spatial representation via cell-level adjustments. Although consistency with basin-scale distribution is used as a convergence constraint, it does not uniquely determine the resulting spatial configuration.

[19] To ensure methodological comparability, a consistent default configuration was adopted for the three categories of input data: the DEM was coarsened using the *Hydro-Aware* method, land cover and soil type data were coarsened using the *Majority* method, and continuous initial condition data were coarsened using the *Mean* method. When evaluating alternative schemes for one category, the other two were kept at their default settings to control confounding effects and maintain structural and functional consistency.

*Table 1. Summary of data-specific coarsening methods.*

| Input Type | Method | Description and Key Features |
|---|---|---|
| **DEM (Continuous)** | *Mean* | Calculates the mean value of fine-resolution cells, producing a smoother representation of the terrain surface. |
| | *Hydro-Aware* | Weights elevation values by flow accumulation to emphasize flow pathways and slope-dominant cells, helping to preserve hydrological connectivity and terrain morphology. |
| **Land cover and Soil type (Categorical)** | *Majority* | Assigns the most frequent class within each coarse grid cell as the representative category. |
| | *Hydro-Aware* | Weights classes by flow accumulation to emphasize land cover and soil types along major drainage pathways, so that coarse-grid distributions better represent key hydrological controls. |
| | *Auto-Weight* | Adaptively optimizes class weights so that aggregated category proportions remain close to their original distribution, helping to reduce distributional bias. |
| | *Auto-Reassign* | Iteratively reallocates classes based on differences between pre- and post-coarsening distributions to improve spatial representativeness and reduce classification distortion. |
| **Continuous Initial Conditions (e.g., soil moisture)** | *Mean* | Calculates the mean value of fine-resolution grid values. |
| | *Hydro-Aware* | Weights values by flow accumulation to emphasize hydrologically dominant areas along main drainage pathways, so that coarse-grid patterns can better represent key hydrological patterns. |
| | *Landcover-Aware* | Averages values within the same land cover type to preserve ecological process consistency. |
| | *Soil-Aware* | Averages values within the same soil type to maintain physical and chemical continuity of soil properties. |

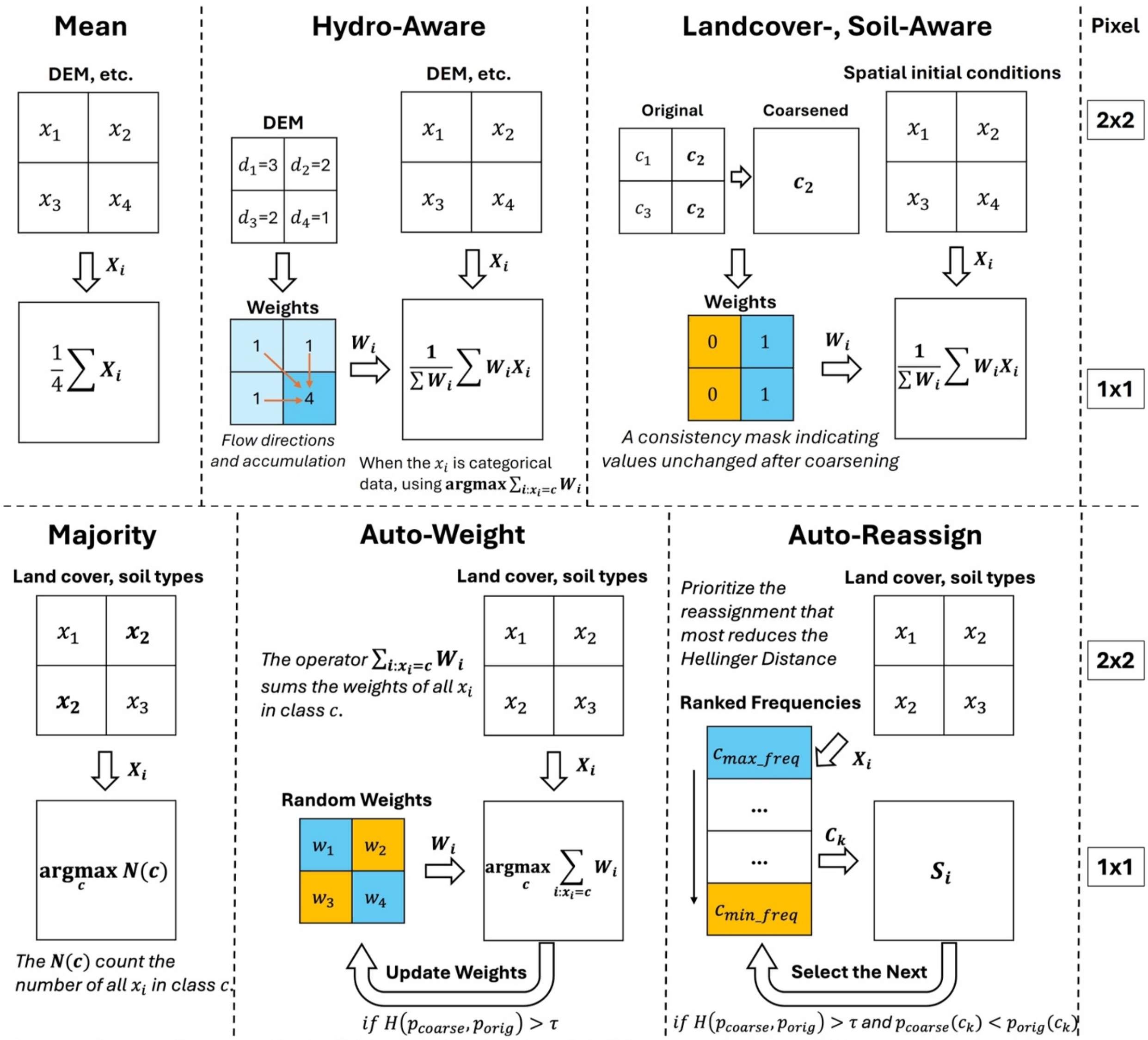

Here, $x$ denotes the value of an individual pixel, $c$ the class label for categorical data, $W$ the associated weight, and $S$ the coarsened class assignment. The operator $\mathbf{argmax}_c(\cdot)$ identifies the class with the largest total weight, $H(\cdot)$ denotes the Hellinger Distance, $\mathbf{p}_{coarse}$ and $\mathbf{p}_{orig}$ represent the coarsened and original class distributions, whereas $\mathbf{p}_{coarse}(\cdot)$ and $\mathbf{p}_{orig}(\cdot)$ refer to the proportion of a specific class, $\tau$ is the threshold.

*Fig. 2 Overview of coarsening strategies for different input types.*
***(a) Mean:*** *Computes the arithmetic average of pixel values (valid only to continuous variables).*
***(b) Hydro-Aware:*** *Aggregates values using flow-direction and flow-accumulation information as weights to preserve hydrological connectivity and flow paths.*
***(c) Landcover- and Soil-Aware:*** *Constrains the coarsening of continuous initial conditions so that the coarse-scale value remains physically consistent with the dominant land cover or soil class in the coarsened area.*
***(d) Majority:*** *Assigns the output pixel to the most frequent class* $c$ *within the coarsened area.*
***(e) Auto-Weight:*** *Determines the output class using a weighted sum. The weights* $w_i$ *are iteratively updated until the difference between the coarse and original class distributions falls below a threshold* $\tau$.
***(f) Auto-Reassign:*** *Selects a class* $c_k$ *from the frequency-ranked list to correct distributional deviations. If the Hellinger Distance* $H$ *exceeds* $\tau$*, the pixel is reassigned to a class for which the coarse-scale proportion* $p_{coarse}$ *is lower than its original proportion* $p_{orig}$.

### 2.3 Coarsening Performance Evaluation Methods

[20] To quantitatively assess similarity between the original high-resolution model and its coarsened counterparts, two complementary aspects were considered: temporal consistency and preservation of spatial distribution. Together, these metrics provide an integrated framework for evaluating model performance across spatial resolutions. To reduce the influence of individual watershed characteristics on the evaluation, the analysis was conducted across 24 sub-basins representing diverse hydro-ecological conditions. In addition, 19 coarsening factors (2-20) were adopted to systematically capture model responses over a continuous range of spatial resolutions, enabling a comprehensive assessment of model stability and performance degradation across scales.

**(1) Temporal Similarity Assessment**

[21] At the watershed outlet, the similarity between time series simulated by the original and coarsened models was evaluated using the mean of the *Nash–Sutcliffe Efficiency* (NSE) and the *Kling–Gupta Efficiency* (KGE) metrics *(Gupta et al., 2009)*. *NSE* measures how closely the coarsened simulations reproduce the temporal dynamics of the original high-resolution simulations (focusing on overall trend fidelity), whereas *KGE* incorporates correlation, bias, and variability components to provide a more comprehensive and balanced assessment of model behavior. In this study, we used ***NSE–KGE***, the arithmetic mean of NSE and KGE, to capture complementary aspects of temporal similarity and statistical consistency.

**(2) Spatial Similarity Assessment**

[22] The *Hellinger Distance* was used to quantify the divergence between pre- and post-coarsening distributions for categorical spatial inputs such as land cover and soil type data *(Beran, 1977; Virdee et al., 2025)*. Defined as a geometric measure between probability distributions, the Hellinger Distance effectively captures changes in the relative proportions of categories, thereby reflecting the degree to which spatial resampling preserves the relative proportions of ecological and pedological categories. Smaller values indicate higher similarity and better retention of spatial structure.

## 3 Results (1750 words)

### 3.1 Effects of DEM Coarsening on Watershed Delineation and Model Performance

[23] Coarsening of DEMs affects not only the smoothness of terrain morphology but also directly determines the fidelity of watershed boundaries and flow connectivity. While the *Mean* method (which assigns each coarse cell the average elevation of the corresponding high-resolution grid) is straightforward to implement, it can easily lead to the loss of fine-scale topographic features and misrepresentation of watershed partitioning. In this study, the *Mean* method is used as a baseline to represent simple spatial aggregation approaches that do not explicitly preserve hydrological topology. To clearly illustrate these structural effects, we present a representative example from the Big Beef subbasin, with similar patterns observed in other basins. Mean-based coarsening caused noticeable boundary shifts, with several flow accumulation cells incorrectly merged into the watershed, leading to substantial changes in basin extent (**Fig. 3(b)**). In contrast, the "*Hydro-Aware method*" incorporated flow accumulation and flow-direction constraints during coarsening, so that the coarsened terrain preserves hydrological connectivity and watershed morphology of the original DEM (**Fig. 3(a)**), thereby effectively reducing structural distortion (**Fig. 3(c)**).

[24] To further evaluate the influence of these structural differences on model performance across all 24 subbasins, the *Mean* and *Hydro-Aware* methods were compared across a range of coarsening factors (from 2 to 20, where $n$ indicates an $n$-fold reduction in resolution). The boxplots in **Figs. 3(d)–(e)** present the NSE–KGE values quantifying the similarity of streamflow and nitrate loss ($NO_3$) simulations between the coarsened models and the reference high-resolution model. Each box shows the distribution of NSE–KGE values across all sub-basins under a given coarsening factor and method, with box bounds representing the 25th and 75th percentiles, the line inside showing the median, and whiskers indicating 1.5 times the interquartile range. For streamflow, the *Hydro-Aware* method consistently maintained high agreement with the reference model across all scales. Its performance not only exceeded that of the *Mean* method, but also declined less sharply with

increasing coarsening, demonstrating superior stability and cross-resolution robustness. Even under the 15× to 20× coarsening, the *Hydro-Aware* method retained an median NSE–KGE above 0.85, whereas the *Mean* method showed large fluctuations, occasionally dropping below 0.3. For $NO_3$ loss, both methods showed an overall decline in similarity with increasing coarsening, and the distributions overlapped substantially across coarsening factors. In contrast to streamflow, the *Hydro-Aware* method did not show a clear and consistent advantage over the mean method (**Fig. 3(e)**).

[25] Note that computational cost decreases approximately with the square of the coarsening factor. For example, when the resolution is reduced by a factor of three, the number of grids in VELMA decreases to about one-ninth of the original, potentially enabling a ninefold speedup. By retaining acceptable accuracy, moderate coarsening can substantially reduce computational burdens and enhance model scalability for large-scale and multi-scenario simulations.

[26] Overall, the *Hydro-Aware* coarsening method balances physical consistency and computational efficiency. By preserving terrain structure and hydrological connectivity while reducing computational demand, it enhances model robustness and scalability across spatial resolutions, providing a consistent terrain foundation for multi-scale ecohydrological simulation.

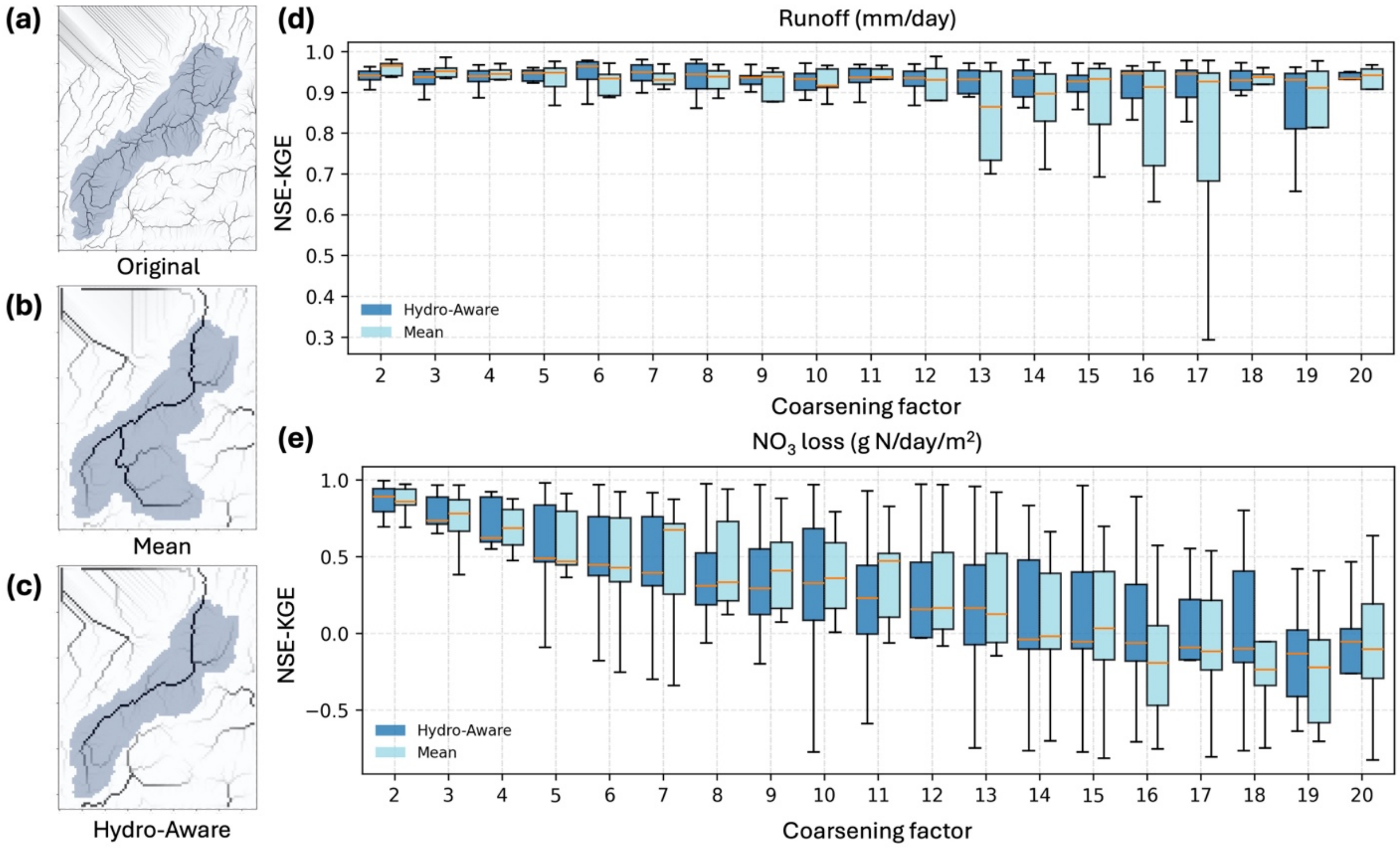

*Fig. 3 Comparison of DEM coarsening methods and their effects on model performance. (a) Original high-resolution DEM of the Big Beef subbasin. (b) Mean-based and (c) Hydro-Aware DEMs coarsened by a factor of 5. (d–e) Boxplots of model performance across coarsening factors (2–20) across the 24 subbasins, evaluated by NSE–KGE for (d) runoff and (e) nitrate loss ($NO_3$); orange lines indicate the median values.*

## 3.2 Effects of Land cover and Soil type Coarsening on Model Performance

[27] When predicting streamflow and nitrogen loss at the watershed-scale, the *Majority* coarsening method has notable limitations for land cover and soil type inputs. This approach implicitly assumes that the dominant class within each coarse grid cell adequately represents the functional behavior of that cell, an assumption that rarely holds in practice. For example, certain soil types with high permeability may exert a dominant influence

on runoff generation, whereas specific land cover types (such as pasture or livestock areas), although spatially limited, can control the spatial distribution and dynamics of nitrogen cycling. Such high-impact but small-area units are easily masked or omitted during the majority aggregation process, leading to a loss of critical distributional information and deviations from high-resolution model behavior.

[28] This issue is quantitatively illustrated using the Samish subbasin as an example, based on the results of majority-based coarsening (**Fig. 4)**. The figure shows the relationships among model similarity (NSE–KGE), the degree of change in input distributions (measured by the Hellinger Distance of land cover and soil type distributions), and the coarsening factor. For both streamflow (**Figs. 4(a)–(b)**) and nitrate loss (**Figs. 4(c)–(d)**), model similarity declines systematically as the deviation in input distributions increases, with stronger degradation observed at higher coarsening levels. This indicates that under the majority-based coarsening scheme, simplification and distortion of input distributions can contribute to cross-resolution deviations and weaken cross-resolution consistency. Similar relationships are consistently observed across other basins. Therefore, preserving realistic and representative input distributions is important for maintaining stable model performance across spatial scales.

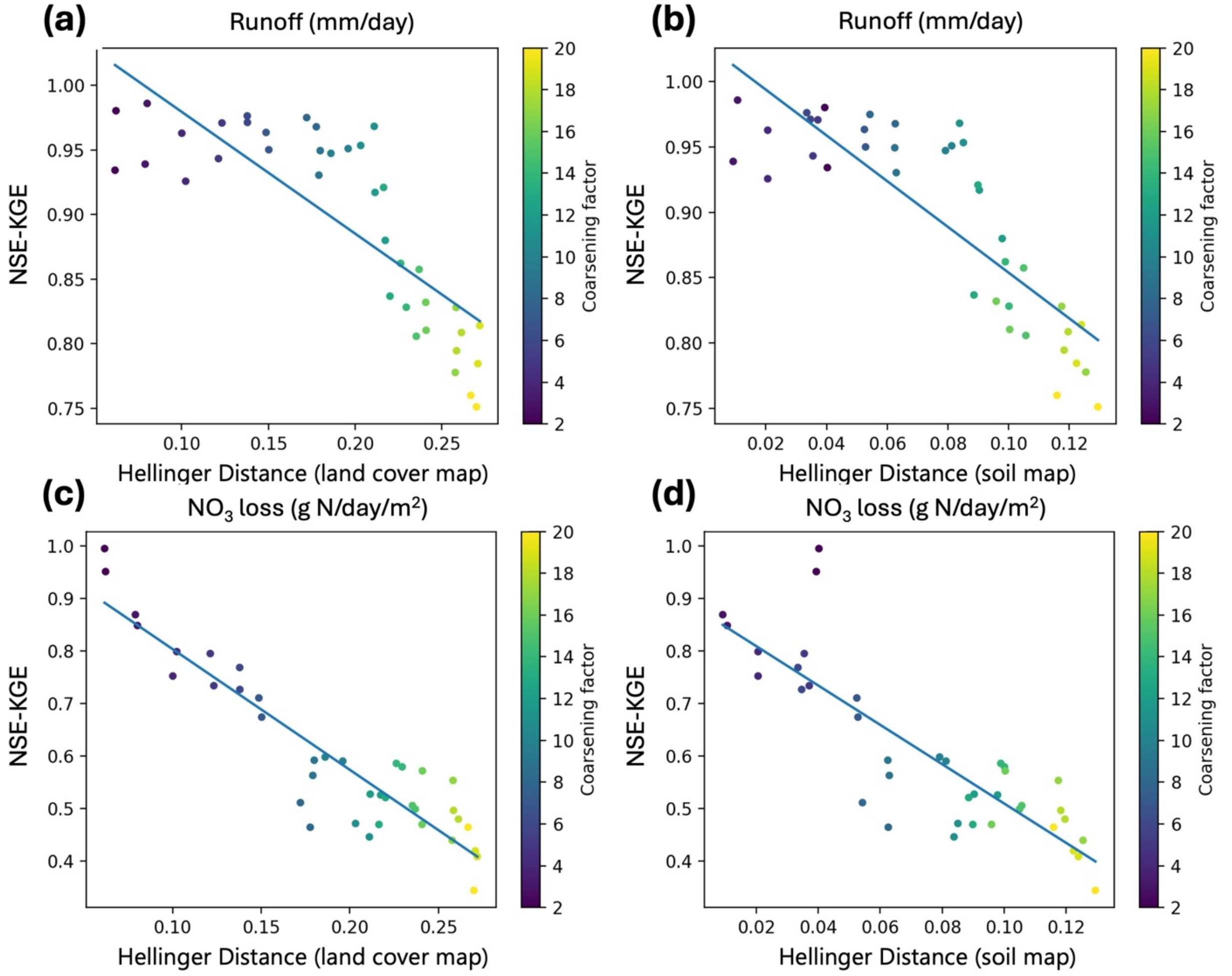

*Fig. 4 Relationships among model similarity, input-distribution divergence, and coarsening factor in the Samish subbasin. (a–b) Relationships between NSE–KGE for runoff and Hellinger distances of land cover and soil type distributions. (c–d) Corresponding relationships for nitrate loss ($NO_3$). Hellinger Distance quantifies the deviation between coarsened and original categorical distributions, and color represents the coarsening factor.*

[29] To address the above issue, this study proposes two improved algorithms: *Auto-Weight* and *Auto-Reassign*. The former adaptively weighs different land cover and soil types during the coarsening process to better preserve that key spatial distribution characteristics remain consistent. The latter employs an automatic reallocation strategy to retain small but influential categories within coarse grid cells. Compared with the *Majority* and *Hydro-Aware* methods, the NSE-KGE results across all subbasins (**Fig. 5**) show that the four approaches perform similarly for runoff simulations (**Fig. 5(a)**), whereas *Auto-Weight* and *Auto-Reassign* tend to maintain higher similarity for nitrate loss ($NO_3$; **Fig. 5(b)**). Overall, model similarity decreases with increasing coarsening factor, particularly for $NO_3$ loss, indicating that nitrogen cycling is more sensitive to spatial resolution change in land cover and soil type inputs. Nevertheless, *Auto-Weight* and *Auto-Reassign* maintain relatively higher NSE–KGE scores at several coarse resolutions, suggesting improved robustness under some coarsening conditions. This improvement is consistently observed across basins, although the magnitude varies depending on watershed characteristics. Runoff simulations are mainly controlled by topography and are therefore less sensitive to the categorical inputs such as land cover and soil type data. In contrast, nitrogen dynamics depend more strongly on the spatial distribution of small but high-impact classes, which the two improved algorithms are designed to better retain.

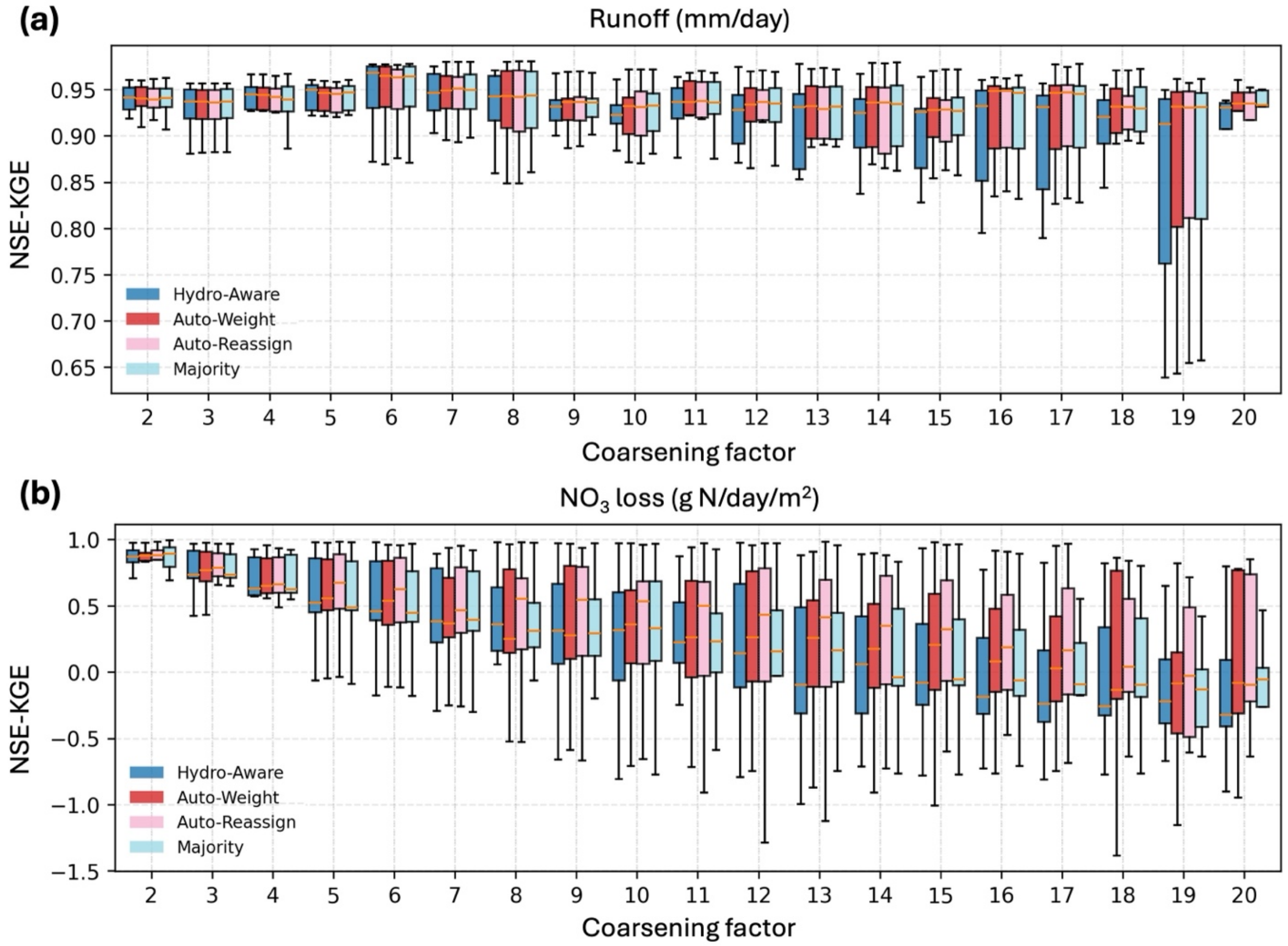

*Fig. 5 Comparison of land cover and soil type coarsening methods and their effects on model performance across all subbasins. (a) Runoff and (b) nitrate loss ($NO_3$) simulations under four coarsening schemes: Majority, Hydro-Aware, Auto-Weight, and Auto-Reassign. Boxplots show performance (NSE–KGE) across coarsening factors (2–20); orange lines denote median values.*

[30] The *Auto-Weight* and *Auto-Reassign* methods generally improve NSE-KGE, although the enhancement is modest. This limited improvement is most evident in basins dominated by Evergreen Forest, where other land

cover types occupy relatively small areas and therefore have less influence on model outputs. For example, in the Dungeness subbasin, Evergreen Forest accounts for 73.1% of the area, while Bare Land and other land-cover types occupy much smaller proportions (**Fig. 6(a)**). As a result, the $NO_3$ simulations produced by the *Auto-Weight* and *Auto-Reassign* are similar to those from the *Majority* method in Dungeness subbasin (**Fig. 6(b)**), and their advantages in preserving land cover distribution are less evident. In contrast, subbasins such as Samish (**Fig. 6(a)**) exhibit more heterogeneous land cover composition, where diverse land types coexist and intensive livestock activities make Pasture areas critical to nitrogen export. In these cases, *Auto-Weight* and *Auto-Reassign* generally outperform the *Majority* method and both better maintain model similarity under large coarsening factors (**Fig. 6(c)**), indicating improved representation of nitrogen dynamics. Similar basin-dependent patterns are also observed in other subbasins with comparable land cover compositions.

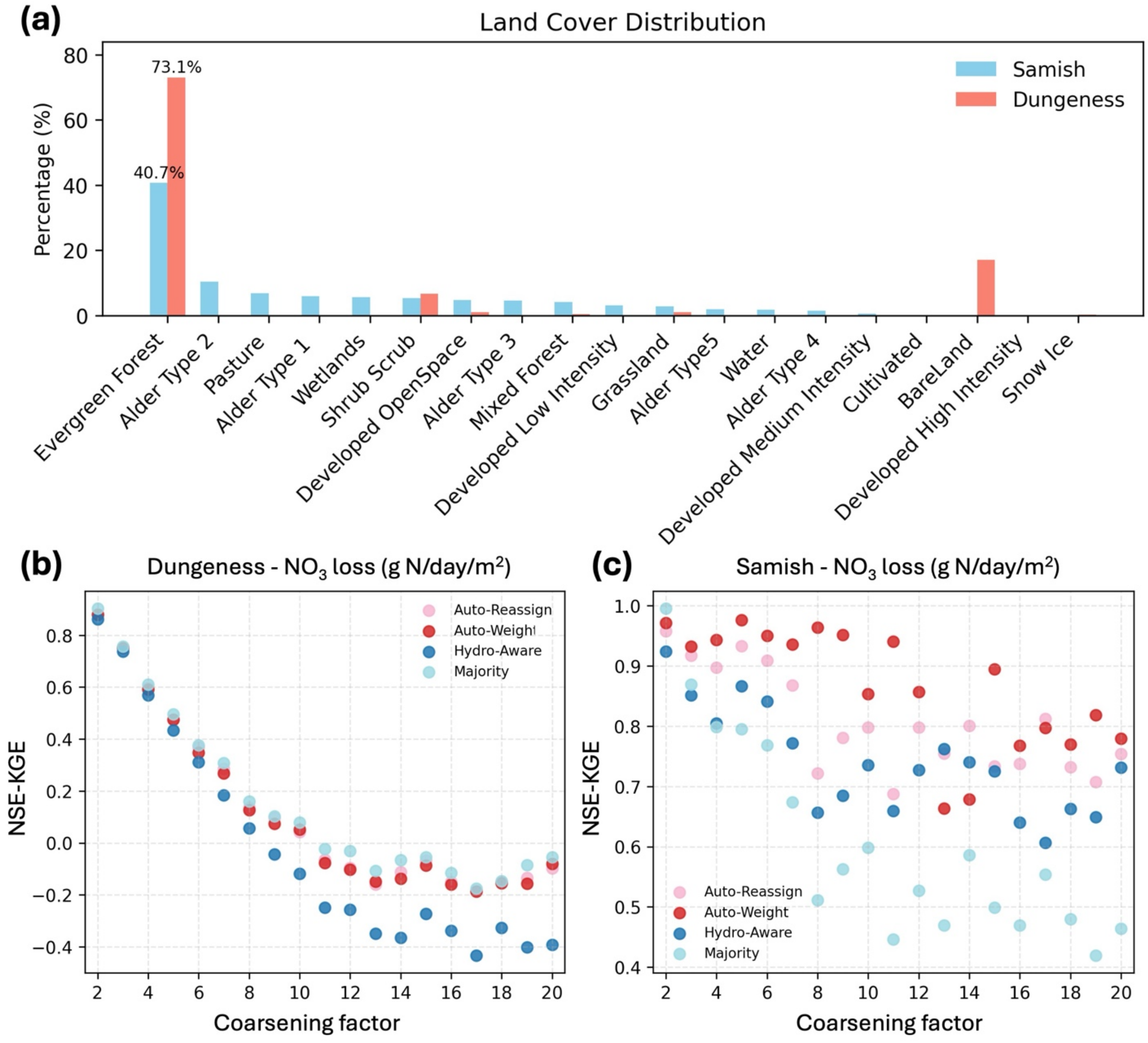

*Fig. 6 Land cover composition and nitrate loss similarity under different categorical coarsening methods. (a) Land cover distributions of the Samish and Dungeness basins. (b–c) NSE–KGE similarity for nitrate loss ($NO_3$) across coarsening factors (2–20) using Majority, Hydro-Aware, Auto-Weight, and Auto-Reassign methods in (b) Dungeness and (c) Samish.*

[31] Overall, while both *Auto-Weight* and *Auto-Reassign* often show higher NSE–KGE similarity than the *Majority* and *Hydro-Aware methods*, differences remain between the two improved algorithms. The *Auto-Weight* applies a weighting scheme to adjust class representation during coarsening, offering good generality and transferability but limited responsiveness to local variations. By contrast, *Auto-Reassign* adaptively reallocates each selected coarse cells based on distributional discrepancies between the coarsened and original maps, providing greater flexibility in retaining small but important categories and local categorical heterogeneity.

### 3.3 Effects of Initial Condition Coarsening on Model Performance and Temporal Stability

[32] Coarsening of spatially distributed initial conditions (e.g., soil moisture and nutrient pools) can also substantially influence model outputs. The most straightforward approach is the *Mean* coarsening method, which averages all grid cells within a coarse unit. However, different land cover and soil types contribute unequally to hydrological and biogeochemical processes, and their corresponding initial states should reflect this variability.

[33] To better represent spatial heterogeneity, three improved initial condition coarsening methods were introduced: *Hydro-Aware*, *Landcover-Aware*, and *Soil-Aware*. The *Hydro-Aware* method weights the original initial conditions according to flow accumulation information to better represent hydrological convergence features. The *Landcover-Aware* and *Soil-Aware* methods apply weighted averaging based on land cover and soil type consistency, respectively, thereby improving the representation of ecological or edaphic characteristics within each coarse unit.

[34] **Fig. 7** indicates that, for runoff simulations (**Fig. 7 (a)**), differences among the four coarsening methods are generally small, and NSE-KGE remains high across coarsening factors, with only modest deterioration at larger coarsening factors. In contrast, nitrate loss ($NO_3$) simulations (**Fig. 7(b)**) are more sensitive to initial condition coarsening: NSE-KGE generally decrease and variability increases with larger coarsening factor rises. Generally, the three improved methods outperform the simple *Mean* approach. Among them, the *Landcover-Aware* method generally shows the highest similarity and comparatively stable performance across coarsening factors, suggesting that land cover based treatment of initial conditions better captures the spatial and ecological controls on nitrogen cycling.

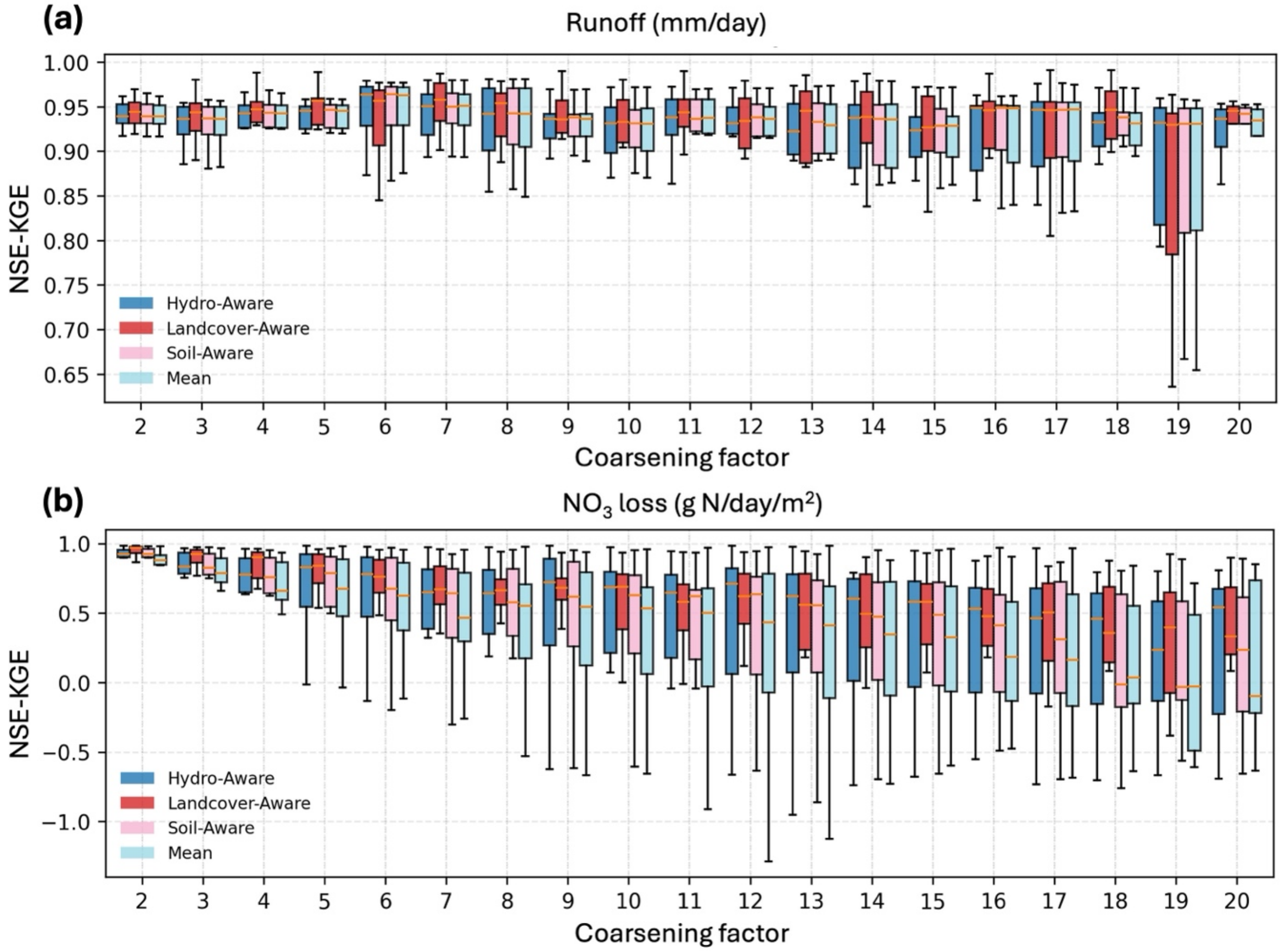

*Fig. 7 Comparison of initial condition coarsening methods and their effects on model similarity across the 24 subbasins. (a) Runoff and (b) nitrate loss ($NO_3$) simulations under four initial condition coarsening schemes: Mean, Hydro-Aware, Landcover-Aware, and Soil-Aware. Boxplots show NSE-KGE similarity between coarsened and reference high-resolution simulations across coarsening factors (2–20); orange lines denote median values.*

[35] To further examine the temporal effects of initial conditions for the Samish subbasin, normalized time series of deviations between the coarsened and original models were analyzed (**Fig. 8**). For runoff simulations, the four coarsening methods produce very similar temporal deviation patterns, with only minor differences among methods. In contrast, nitrate loss simulations exhibit more pronounced deviations, particularly during the early simulation period. The *Landcover-Aware* method generally produces narrower deviations during parts of this initial phase, suggesting better preservation of spatial controls on nitrogen dynamics. As the simulation proceeds, especially after approximately four to five years, the differences gradually diminish, indicating reduced sensitivity to initial condition differences as the model states adjust through internal feedback processes.

[36] This behavior indicates that while initial conditions can introduce significant deviations during the early phase of simulation, but their influence tends to decay over time as the model adjusts internally. For example, hydrological processes such as infiltration, runoff generation, and evapotranspiration rapidly redistribute water within the system, steering the model toward a climate-forced equilibrium and progressively reducing the influence of initial differences. Faster-responding variables, such as runoff and soil moisture, therefore approach stable behavior relatively quickly. In contrast, slower processes such as nitrogen cycling remain more sensitive to initial condition coarsening, making appropriate coarsening schemes, particularly the *Landcover-*

*Aware* approach, and sufficiently long spin-up periods important for stable and reliable model outputs. This behavior is consistently observed across basins, although the time required to reach equilibrium may vary depending on watershed characteristics.

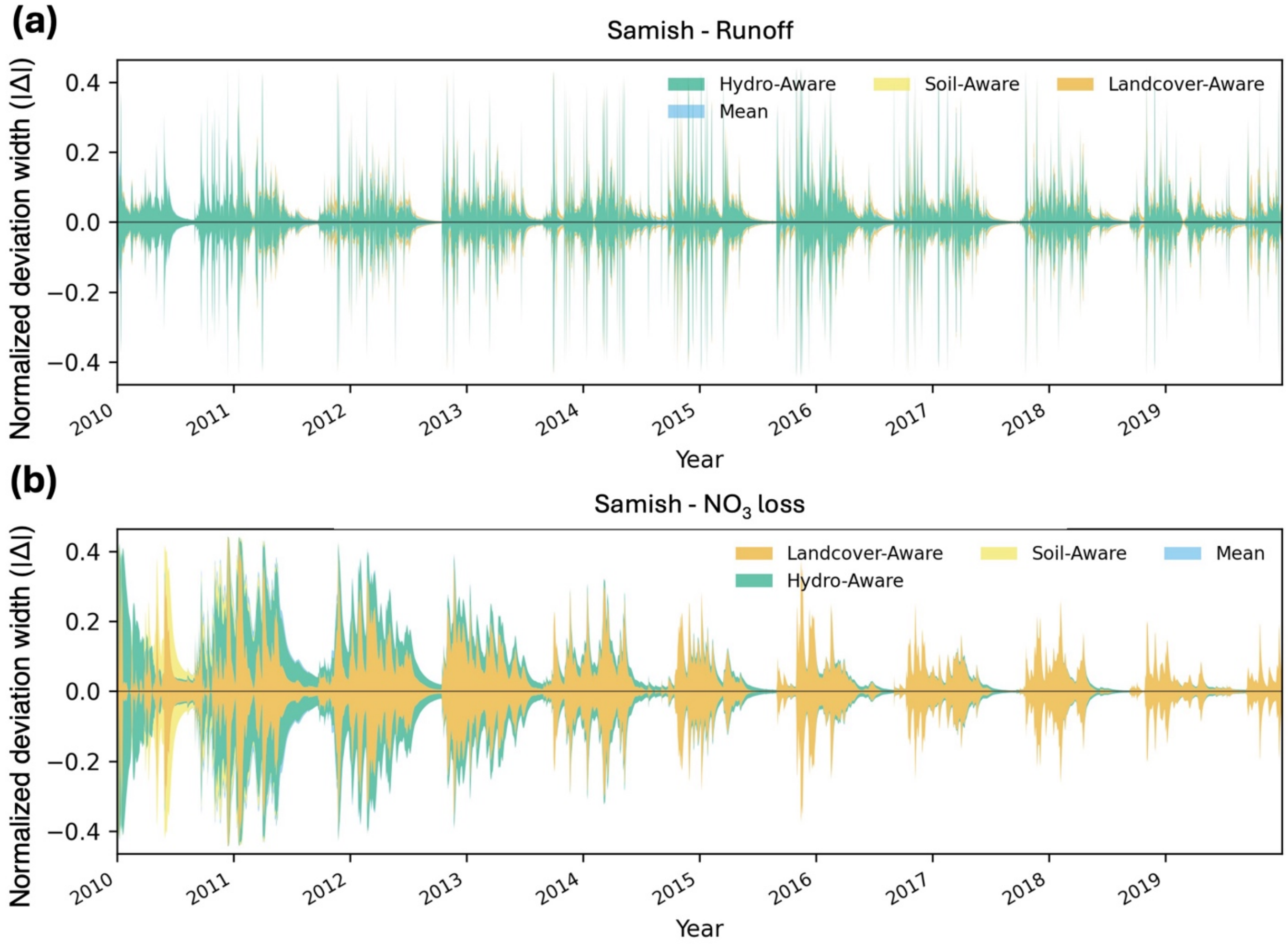

*Fig. 8 Temporal evolution of normalized model deviations under different initial condition coarsening methods in the Samish subbasin. (a) Runoff and (b) nitrate loss ($NO_3$) simulated using four initial condition coarsening schemes: Mean, Hydro-Aware, Landcover-Aware, and Soil-Aware. Shaded ribbons show the absolute normalized deviations (|Δ|) between coarsened simulations and the reference high-resolution simulation over time; narrower ribbons indicate smaller deviations.*

## 4 Discussion (1824 words)

### 4.1 Structure–Function Coherent Coarsening Framework for Cross-Scale Modeling

[37] Rather than treating coarsening as a purely geometric or statistical resampling problem, this study demonstrates that preserving process-relevant structure and function is useful for achieving cross-scale consistency. In this context, the *Hydro-Aware* method better preserves watershed hydrological topology during DEM coarsening, helping maintain the integrity of flow networks across spatial scales. This improved structural consistency leads to more stable streamflow simulations and may also reduce deviations in nitrogen export simulations. The findings align with previous studies that emphasize the critical role of topographic connectivity in shaping hydrological responses *(Epting et al., 2018; Jencso et al., 2009; Ortíz-Rodríguez et al., 2022)*, indicating that structural consistency represents not only an extension of geomorphic form but also an important condition for process robustness and dynamic system stability. In contrast, traditional *Mean* coarsening methods often neglect topographic constraints, leading to flow path distortions, loss of fine-scale

terrain features, and restructured runoff patterns, thereby potentially introducing systematic bias *(Akbar et al., 2024; Almeida et al., 2024)*. By incorporating flow-direction and flow-accumulation constraints during coarsening, the *Hydro-Aware* approach can effectively reduce structural distortions and stabilizes model performance across resolutions, providing a solid physical foundation for cross-scale simulation *(Minh et al., 2024; Rasheed et al., 2024; Rocha et al., 2023)*.

[38] Analyses of land cover and soil types further reveal that function preservation is an important complement to structural consistency. The commonly used *Majority* method implicitly assumes that areal proportion equates to process contribution, an assumption that fails in highly heterogeneous basins. Small but process-dominant units (e.g., highly permeable sandy soils or intensive pasture lands) are often masked during coarsening, leading to biased nitrogen-cycle and water-quality simulations *(Horta et al., 2024; Rumbold et al., 2023; Zhang et al., 2023)*. The *Auto-Weight* and *Auto-Reassign* methods employ adaptive weighting and dynamic redistribution to enhance the representation of high-contribution areas within coarsened units, helping mitigate the "*dominant-class bias*" inherent in majority-based aggregation. These results suggest that process consistency depends on the functional weighting of inputs rather than their areal dominance, highlighting the coupled roles of structural integrity and functional stability. Together, these mechanisms sustain system robustness at the ecological–functional level, enabling models to maintain high similarity to the reference high-resolution simulations and interpretability even under aggressive coarsening. This concept aligns with recent advances in multiscale structure–process coupling *(Arnold et al., 2023; Polcher et al., 2023; Yin et al., 2023)*, offering a new perspective for process-consistent cross-scale modeling.

[39] Analysis of initial conditions further indicates that model consistency depends not only on spatial structure but also on temporal adjustment. Hydrological variables, such as streamflow and soil moisture, respond rapidly to initial condition differences, whereas slower biogeochemical processes, such as nitrogen cycling, exhibit pronounced temporal lags *(Pang et al., 2022; Wu et al., 2022)*. For example, in the Samish subbasin, model outputs show larger deviations during the initial simulation years, but these deviations generally diminish after approximately four to five years. This observation is consistent with previous findings *(Dwivedi et al., 2025; Sidle, 2021; Xia et al., 2023)*, and also suggests that structural consistency and functional preservation jointly help reduce transient deviations and support stable long-term model behavior.

[40] Finally, this study revisits the implicit assumption that "*spatial similarity necessarily implies response similarity.*" Unlike approaches that treat spatial coarsening as a purely geometric aggregation, this work presents and validates two complementary mechanisms: *Structure-Consistent* and *Function-Preserving* coarsening. The former, represented by the *Hydro-Aware* method, emphasizes maintaining topographic–hydrological integrity and reducing distortions in flow connectivity, while the latter, represented by the *Auto-Weight* and *Auto-Reassign* methods, helps preserve the influence of small, yet process-dominant, units during coarsening. Together, they form the *Structure–Function Coherent Coarsening* (SFCC) framework, which simultaneously reduces structural distortions and process-response deviations during spatial coarsening, supporting model stability and robustness across resolutions. Compared with existing frameworks such as *Multiscale Parameter Regionalization* (MPR) and SUMMA, SFCC focuses on preserving structural and functional information at the data-input level, providing a physically consistent and functionally representative pathway for cross-scale ecohydrological modeling.

### 4.2 Applicability, Limitation, and Future Directions

[41] Beyond its theoretical contributions, the SFCC framework demonstrates practical applicability within the VELMA modeling context. The method has been successfully applied to multiple subbasins of the Salish Sea using the VELMA model to support efficient cross-resolution simulation and calibration. It can substantially reduces computational costs and runtime, thereby improving calibration efficiency by reducing the cost of repeated model evaluations. These results establish an efficient and replicable pathway for large-scale ecohydrological modeling and watershed management. In practical applications, direct coarsening from the finest resolution is generally preferred, as it minimizes cumulative information loss and better preserves structural and functional consistency across scales. Progressive coarsening from intermediate resolutions

provides a flexible alternative, but its applicability depends on the sensitivity of target variables to spatial information loss and the smoothness of degradation across scales. It is more suitable for hydrological responses, such as streamflow, which are relatively robust to spatial aggregation, whereas for processes sensitive to fine-scale heterogeneity, such as nitrate dynamics, stepwise aggregation may amplify accumulated errors. Therefore, progressive coarsening should be applied with caution.

[42] Moreover, SFCC can be adapted to other diverse process-based ecohydrological models (e.g., SWAT, RHESSys, and DHSVM), maintaining structural consistency among terrain, hydrology, and ecosystem components. Such integration may reduce spatial aggregation bias and support more efficient calibration *(Thébault et al., 2024; Tudaji et al., 2025; Zhang et al., 2025)*, forming a structural foundation for cross-model comparison, parameter transfer, and multi-basin simulation. SFCC also shows strong potential for integration with emerging variable-resolution ecohydrological modeling frameworks *(Galelli et al., 2025)*. While variable-resolution approaches aim to improve efficiency through heterogeneous grid configurations, SFCC provides a structure-preserving aggregation strategy that maintains hydrological connectivity and land–soil functional heterogeneity during scale transitions. This enables more consistent coupling between fine- and coarse-resolution units, supports cross-scale parameter consistency, and offers guidance for adaptive grid design in spatially heterogeneous regions.

[43] At the regional and policy levels, the SFCC framework provides a useful foundation for multiscale scenario analysis and integrated watershed management. In regions characterized by complex terrain and high process heterogeneity, SFCC can improve model consistency across resolutions, enabling scenario differences to reflect more likely to reflect process responses rather than artifacts of input resolution *(Yang et al., 2022)*. It also helps to mitigate integration biases that often arise when combining multi-source datasets with inconsistent spatial resolutions, allowing for more coherent representation of watershed processes across scales. This capability is valueable for nitrogen-load reduction, ecological restoration planning, and climate adaptation assessments *(Liu et al., 2024; Wu et al., 2023)*. Furthermore, by reducing information loss during spatial coarsening, the function-preserving mechanism enhances computational efficiency while maintaining acceptable cross-resolution consistency, supporting regional and cross-basin simulations and decision analyses *(Chen and Cai, 2024; Tapas et al., 2025)*.

[44] In addition, SFCC establishes a useful physical basis for cross-resolution calibration and parameter transfer. By simultaneously maintaining structural and functional consistency, models of varying resolutions may become more consistent in their parameter interpretation, potentially improving calibration efficiency and parameter transferability. This perspective is consistent with recent progress in multiscale parameter regionalization and model harmonization research *(Droppers et al., 2024; Feigl et al., 2022; Schweppe et al., 2022)*, highlighting that dual structural–functional constraints may support parameter transferability and model generalization. Consequently, SFCC exhibits broad potential for multi-model integration, surrogate modeling, and regional composite simulation, providing a methodological basis for multiscale ecohydrological modeling systems.

[45] Despite the observed advantages of SFCC in enhancing cross-resolution consistency, this study has several limitations. First, the study region is concentrated in the Salish Sea basin, and its temperate maritime climate limits direction evaluation of the method's applicability in arid or alpine environments. Second, the employed VELMA model may simplifies slow-changing processes such as groundwater–surface water exchange, subsurface denitrification, and long-term carbon–nitrogen coupling, potentially underestimating the influence of slow variables on system evolution *(Irvine et al., 2024; Shin et al., 2024)*. Third, the simulation period of roughly ten years may be insufficient to capture the nonlinear cumulative impacts of climate variability and extreme events over multi-decadal timescales

[46] Future research could advance along three main directions:

- **Multi-source data fusion and algorithm optimization**: Integrating high-resolution LiDAR DEMs, SoilGrids, NDVI, and dynamic land cover data to develop more *process-aware* coarsening methods that jointly optimize terrain, land, and soil information *(Aristizabal et al., 2024; Huang et al., 2025)*.

- **Graph-based coarsening**: Employing graph-clustering and community detection algorithms (e.g., spectral clustering and Louvain community detection) to achieve adaptive spatial simplification while better preserving hydrological topology, offering a structured pathway for efficient modeling of complex basins *(Chen et al., 2023; Cole et al., 2023)*.
- **Machine-learning based mapping and surrogate modeling**: Exploring the compatibility between SFCC and surrogate learning approaches, such as Long Short Term Memory networks (LSTMs) and Graph Neural Networks (GNNs), to construct stable input–output mappings and enhance model transferability and generalization *(Farahmand et al., 2023)*.

[47] By integrating multi-source data, graph-structured representations, and machine-learning mechanisms, future ecohydrological modeling is expected to move beyond resolution-dependent paradigms toward more transferable, scalable, and adaptive modeling systems. Such advances would provide a stronger scientific basis and decision-support capacity for regional water-resource management and sustainable ecosystem governance across multiple scales and scenarios.

## 5 Conclusion (292 words)

[48] This study revisits the long-standing scaling issue in ecohydrological modeling from a new perspective, aiming to reduce input resolution while better preserving hydrological structure and process-relevant information. We propose a *Structure–Function Coherent Coarsening* (SFCC) framework that integrates a structure-consistent pathway, represented by a *Hydro-Aware* DEM coarsening method, with a function-preserving pathway, represented by *Auto-Weight* and *Auto-Reassign* algorithms for land cover and soil data. Modeling results show that acting directly on the input layers, SFCC better preserves drainage topology, spatial representativeness, and process relevance, providing a simple and effective method that stabilizes model behavior across resolutions.

[49] Three main conclusions emerge. First, structural consistency, rather than geometric similarity, is essential for cross-scale similarity, so that maintaining hydrological topology during DEM coarsening is crucial. Second, structural preservation alone can be insufficient when categorical inputs drive biogeochemical processes, and preserving the contribution of high-impact classes is important, particularly in heterogeneous basins where nitrogen dynamics are governed by small pasture or permeable soil patches. Third, structural consistency and functional preservation work together to reduce transient deviations and support long-term system stability.

[50] Accordingly, the SFCC framework can provides a physically grounded solution that reconciles computational efficiency with process fidelity. It can support more efficient calibration at coarser resolutions while maintaining parameter transferability across scales, facilitate the coherent integration of multi-source datasets with differing spatial resolutions, and facilitate multiscale scenario analysis, cross-basin comparison, and integrated watershed management. Future research should integrate SFCC with high-resolution multi-source datasets, explore graph-based and community detection coarsening method, and hierarchical pooling, and couple with LSTM or GNN machine learning models to develop cross-scale modeling systems. These advances could help ecohydrological models move beyond resolution dependence and enhance their capacity to support sustainable water and ecosystem management.

## 6 Acknowledgements

This work was supported by the Allen Family Philanthropies. We thank Jonathan Halama for his valuable assistance and support during this work.

## 7 Open Research

The datasets and software used in this study are publicly available from the sources cited below. Elevation data at 30 m resolution were obtained from the National Elevation Dataset *(U.S. Geological Survey, 2019)*. Land cover data at 30 m resolution were obtained from the National Land Cover Database *(Dewitz, 2021)*. Forest composition and forest age data were obtained from the LEMMA project *(Landscape Ecology Modeling, Mapping and Analysis, 2017)*. Soil restrictive depth and soil organic carbon data were obtained from the SOLUS database *(Soil Survey Staff, 2024)*. Daily precipitation and temperature data were obtained from an 800 m resolution gridded meteorological dataset *(Daly et al., 2021)*. Daily streamflow observations (2009–2019) were obtained from USGS monitoring stations *(U.S. Geological Survey, 2021)*. Monthly water quality observations, including ammonia, nitrate, dissolved organic carbon, and water temperature (2009–2019), were obtained from the EIM database *(Washington State Department of Ecology, 2025)*. Parcel-scale septic system data used in this study are not publicly available because of data-use restrictions, although census block-scale septic system density data are publicly available *(Peterson et al., 2025)*. Digital elevation preprocessing, including dredging and flat processing to enforce valid flow paths, was conducted using JPDEM *(U.S. Environmental Protection Agency, 2017)*. Ecohydrological simulations were performed using VELMA *(U.S. Environmental Protection Agency, 2024)*. All key algorithms and Python scripts supporting spatial coarsening, data processing, model analysis, and figure generation are publicly available in the GitHub repository *(Jiang, 2025)*.

## 8 Conflict of Interest

The authors declare there are no conflicts of interest for this manuscript.